\documentstyle[11pt,epsfig]{article}
\begin{document}
\title{On the range of validity \\ of the QCD-improved parton model}

   \author{A. Szczurek$^{1,\protect
\footnote{E-mail:szczurek@alf.ifj.edu.pl}}$ and V. Uleshchenko$^{1,2}$  \\
   {\it $^{1}$ Institute of Nuclear Physics, PL-31-342 Cracow, Poland  } \\
   {\it $^{2}$ Institute for Nuclear Research, 252028 Kiev, Ukraine  } \\   }
\maketitle
\begin{abstract}
Based on the world DIS data we extract the experimental $F_2^p-F_2^n$
as a function of Bjorken-$x$ and photon virtuality $Q^2$,
using two different methods.
Both methods lead to identical results. We find that the standard PDFs fail
to describe the experimental data below $Q^2 < $ 7 GeV$^2$, which is much
higher than for $F_2^p$ and $F_2^d$ separately. The difference between PDFs
and the experimental data cannot be understood as due to nuclear effects in
the deuteron, and evidently suggests substantial nonsinglet higher-twist effects.
The trend of the experimental data is approximately explained by a recent
two-component model of the nucleon structure functions and suggests strong
$Q^2$-dependence of the Gottfried Sum, in disagreement with the parton model
interpretation. The large negative
higher-twist effects can explain the difference between the value of
the Gottfied integral obtained recently by the E866 Drell-Yan experiment
at Fermilab and an older NMC result.
\end{abstract}

\vspace{0.5cm}

%------------------------------------------------------------------------

\quad
The QCD-improved parton model (IPM) applies at large virtuality of
the probe. However, the exact range of its applicability is not obvious.
Generally a good desciption of $F_2^p$ and $F_2^d$ within IPM
can be obtained already above $Q^2 \sim $ 1-2 GeV$^2$
\cite{GRV94,MSRT98,CTEQ5}.

In the last decade the precision of experimental data on structure functions
became so good that one can try to obtain the information on
effects beyond the leading twists \cite{VM92,NMC_ht}.

A recent lattice QCD calculation has found an unexpectedly large twist-4
contribution to the structure functions \cite{lattice}.
This is rather difficult to reconcile in the light of the success of
the improved parton model which is known to work empirically down to
$Q^2 \sim$ 1-2 GeV$^2$.

In a recent work\cite{SU1} we have shown that both
the proton and deuteron structure functions can be well explained in
a broad range of $x$ and $Q^2$ also in a two-component model of
the structure function
\begin{equation}
F_2^{p/n}(x,Q^2) = F_2^{p/n,had}(x,Q^2) + F_2^{p/n,part}(x,Q^2) \; ,
\label{F2_decomposition}
\end{equation}
which gives better agreement with data below $Q^2 \sim$ 3 GeV$^2$.
Our model fulfills $F_2^{p/n}(x,Q^2) \rightarrow$ 0
when $Q^2 \rightarrow$ 0, as required by current conservation.
While $F_2^{p/n,had}(x,Q^2) \rightarrow$ 0 by construction, the vanishing
of $F_2^{p/n,part}(x,Q^2)$ is achieved as in the low-x
Bade\l ek-Kwieci\'nski model \cite{BK_model}
\begin{equation}
F_2^{p/n,part}(x,Q^2) =
\frac{Q^2}{Q^2 + Q_0^2} \cdot F_2^{p/n,asymp}(\bar x,\bar Q^2) \; ,
\label{F2_part}
\end{equation}
where $F_2^{p/n,asymp}$ is the standard parton-model structure function
and $\bar x$, $Q^2$ are defined as in \cite{BK_model}.
This simple phenomenological form insures a correct $Q^2 \rightarrow$ 0
limit and is justified by the dispersion method \cite{BK_model}.
For not too small $x$, i.e. in the region we are interested in here,
the modification of the structure function arguments
$x \rightarrow \bar x$, $Q^2 \rightarrow \bar Q^2$ is not needed \cite{SU1}.

Except for a better description of the $F_2$ data in the low-$Q^2$ region
the model from \cite{SU1} has some interesting consequences.
The higher-twist contribution to the structure function can be defined
by
\begin{equation}
\delta^{HT} F_2^{p/n}(x,Q^2) \equiv
F_2^{p/n}(x,Q^2) - F_2^{p/n,asymp}(x,Q^2) \; .
\label{HT_def}
\end{equation}
In the model for $F_2^p$, $F_2^n$ proposed in \cite{SU1} there is a
significant cancellation between positive (VDM-type) and negative (due to
a modification of the partonic component) higher twist contributions
\begin{center}
$\delta^{HT} F_2^{p/n}(x,Q^2) =$
\end{center}
\begin{eqnarray}
&=&F_2^{p/n,had}(x,Q^2) +
\frac{Q^2}{Q^2 + Q_0^2} \cdot F_2^{p/n,asymp}(\bar x,\bar Q^2)
 - F_2^{p/n,asymp}(x,Q^2) \approx \nonumber \\
&\approx& F_2^{p/n,had}(x,Q^2) -
\frac{Q_0^2}{Q^2 + Q_0^2} \cdot F_2^{p/n,asymp}(x,Q^2)
\label{HT_cancel}
\end{eqnarray}
in the range of small and intermediate $x$.

Neglecting a small $\rho$-$\omega$ nondiagonal term due to the
exchange of a Regge trajectory $a_2$
\begin{equation}
\delta^{(\rho-\omega)} F_2^{p/n}(x,Q^2) =
\pm \frac{Q^2}{\pi} \cdot C_{\rho,\omega}^{a_2} \cdot
\frac{m_{\rho}^2}{(Q^2 + m_{\rho}^2)} \cdot
\frac{m_{\omega}^2}{(Q^2 + m_{\omega}^2)}
\cdot \Omega_{\rho,\omega}(x,Q^2) \; ,
\label{nondiagonal}
\end{equation}
we have
%
%%\center{$F_2^{p,had}(x,Q^2) = F_2^{n,had}(x,Q^2) $}
\begin{center}
$F_2^{p,had}(x,Q^2) = F_2^{n,had}(x,Q^2) $
\end{center}
\begin{equation}
\equiv F_2^{had}(x,Q^2) =
\frac{Q^2}{\pi} \sum_V C_V \frac{M_V^4}{(Q^2 + M_V^2)^2}
\cdot \Omega_V(x,Q^2) \; ,
\label{had_def}
\end{equation}
where $C_V$ is a normalization factor which can be related to
the electromagnetic $V \rightarrow e^+ e^-$ decay width and
$\Omega(x,Q^2)$ is a form factor which phenomenologically
accounts for the finite lifetime of the hadronic fluctuation
of the photon \cite{SU1}. Therefore
\begin{eqnarray}
D_G(x,Q^2) &\equiv& F_2^p(x,Q^2) - F_2^n(x,Q^2) =
\nonumber \\
& = &
\frac{Q^2}{Q^2 + Q_0^2} \cdot
\{ F_2^{p,asymp}(x, Q^2) - F_2^{n,asymp}(x, Q^2) \} \;
\label{diff_part}
\end{eqnarray}
for the range of Bjorken-$x$ we are interested in.
Unlike the case of the structure functions there is no cancellation
of hadronic and partonic higher twists. A strong $Q^2$-dependence
of $D_G$ is obvious.
The factor in front of the r.h.s. of Eq.(\ref{diff_part}) assures
also the vanishing of the Gottfried integral for $Q^2 \rightarrow$ 0.
This result is in contrast
with the result obtained by Ball and Forte \cite{BF94} who found
a restoration of the Gottfried Sum Rule
$S_G = \int_0^1 \frac{dx}{x} \{ F_2^p(x,Q^2) - F_2^n(x,Q^2) \}
\rightarrow \frac{1}{3}$ when $Q^2 \rightarrow$ 0.

In order to verify the anticipated effect discussed above let us come to
the status of experimental data on $F_2^p-F_2^n$.
Because $F_2^n$ is not a directly measurable quantity one is forced to
use rather deuteron data.
The simplest approximation, neglecting all nuclear effects in the deuteron,
would be to replace $D_G \equiv F_2^p-F_2^n$ by $F_2^p-(2F_2^d-F_2^p)$.
This direct method is, however, not very useful as far as the existing data
are concerned. There the errors (both systematical and statistical)
for $F_2^p$ and $F_2^d$ are independent and the resulting errors for $D_G$
are too large to reveal the anticipated $Q^2$-dependence.
Instead of the direct subtraction of $F_2^d$ and $F_2^p$
one can use a method proposed by NMC at CERN \cite{NMC}.
In the following two methods will be used to extract $D_G$:
\begin{itemize}
\item (A):
$F_2^p(x,Q^2) - F_2^n(x,Q^2) = F_2^p(x,Q^2) \cdot
\left( 1 - \frac{F_2^n(x,Q^2)}{F_2^p(x,Q^2)} \right)  \; ,$
\item (B):
$F_2^p(x,Q^2) - F_2^n(x,Q^2) = 2 F_2^d(x,Q^2) \cdot
\frac{1 - \frac{F_2^n(x,Q^2)}{F_2^p(x,Q^2)}}
{1 + \frac{F_2^n(x,Q^2)}{F_2^p(x,Q^2)}}  \; ,$
\end{itemize}
where $F_2^n/F_2^p$ is extracted from the measured $F_2^d/F_2^p$ data.
The latter was already used by NMC in the evaluation of the Gottfried
integral in an unpublished analysis of the $Q^2$ dependence of the NMC
data \cite{AB_habilitation}. The two methods above are of course equivalent
when nuclear effects are neglected.
They are equivalent also in a less obvious
case when nuclear effects are taken into account and
$F_2^n/F_2^p$ is replaced by $2 F_2^d/F_2^p - 1$
(this will be assumed hereafter). It can be shown that then both methods
include nuclear effects \footnote{the latter will be marked by a star
in Fig.2 and 3.}
in exactly the same way, identical as the difference
$F_2^p - (2F_2^d - F_2^p)$.

Both methods require knowledge of the ratio
$F_2^n / F_2^p$ for the same $x$ and $Q^2$ points as the structure
functions $F_2^p$ or $F_2^d$ which is not the case for existing experimental
data. For this reason use of a parametrization for $F_2^n/F_2^p$ appears
necessary.
In our analysis we have used a parametrization from \cite{NMC_ht}.
In Fig.1 (top panel) we compare this parametrization with precise
$F_2^d/F_2^p$ data from NMC \cite{NMC_97}.
The dashed lines indicate a 2 \% uncertainty.
%Let us justify our choice.
Below we shall quantify the quality of the parametrization and justify the
choice of its uncertainty band.
In the bottom-left panel we present a percentage of points
above (solid line) and below (dashed line) a trial (rescaled) parametrization
for the ratio, as a function of the relative deviation from the original
parametrization.
The result clearly indicates that statistically the same
amount of points lies above and below the nominal curve.
In the bottom-right panel we show the fraction of NMC data points
in a band of uncertainty as a function of the band width (dashed line).
About 3/4 of the experimental points remain in the 2 \% band of uncertainty.
Because the functional form of the ratio is not known,
and some fluctuations of the ratio are not excluded a priori,
we have also performed the following analysis.
We assign to each experimental point a Gaussian distribution of probability,
centered on the experimental point, with a standard deviation equal
to the experimental error. We then determine the fraction of the area
that lies below such a probability distribution within the parametrization
uncertainty.
We show this quantity averaged over 260 NMC experimental points \cite{NMC_97}
in the bottom-right panel by the solid line marked by $<P>$.
About 60 \% of so-defined probability remain in the 2 \% band of uncertainty.
\footnote{Please note that within standard statistical analysis, assuming
the Gaussian probability distribution, the required fraction
for the standard deviation is 68.2 \%}.
This analysis assumes inherently independence of successive
experimental points and allows fluctuations around
the smooth parametrization.
Because we do not expect neither a full independence of neighbouring
experimental points nor sharp fluctuations of the ratio, the 2 \% uncertainty
band seems to be rather an upper limit for the uncertainty
on $F_2^n / F_2^p$. Taking into account uncertainty of $F_2^n/F_2^p$
at its upper limit should allow us to prove our
thesis about the IPM breaking with a bigger degree of reliability.
We have checked that in the interesting kinematical range the used
parametrization is consistent within experimental uncertainties with
the E665 \cite{E665_ratio} data for $F_2^d/F_2^p$.

The world data on $F_2^p$ and $F_2^d$ from the compilation \cite{F2_data}
was taken in the present analysis.

In Fig.2 we show experimental data for $D_G(x)$
obtained with the two methods for three different values of photon
virtuality $Q^2$ = 1.1, 3.4, 7.0 GeV$^2$ as a function of Bjorken-$x$.
The error bars take into account the inherent uncertainties of
$F_2^p$ (method A) or $F_2^d$ (method B) as well as the uncertainty of
$F_2^n/F_2^p$.
The two bands below each distribution separately show the errors due to the
mentioned above
2\% uncertainty in $F_2^n/F_2^p$ and due to the uncertainty of
$F_2^p$ (lhs) or $F_2^d$ (rhs).
%Shown separately is a band due to the mentioned
%above
%2\% uncertainty in $F_2^n/F_2^p$ and a band due to the uncertainty of
%$F_2^p$ (lhs) or $F_2^d$ (rhs).
As seen from the figure the error of $D_G$ due to the uncertainty of the
$F_2^n/F_2^p$ ratio is substantially bigger than the one due to
the uncertainty of either $F_2^p$ or $F_2^d$. Within
experimental errors the two methods lead to identical results
despite the different data sets used.
We have also checked that the result obtained by
a direct subtraction of the deuteron and proton structure functions
leads
to a consistent result although the scatter of the data is much larger.
Therefore the data obtained in this way can be reliably used to study
the $Q^2$ dependence of $F_2^p-F_2^n$ and/or corresponding
higher-twist effects.

For comparison we show theoretical results obtained with
GRV94 \cite{GRV94} (short-dashed line),
MRST \cite{MSRT98} (long-dashed line) and a recent
CTEQ5 \cite{CTEQ5} (solid line) NLO DIS-scheme parton distributions.
\footnote{The MRST result is shown only above $Q^2>$ 1.25 GeV$^2$
and CTEQ5 above $Q^2>$ 1 GeV$^2$, i.e. in their applicability range.}
Above $Q^2 >$ 7 GeV$^2$ the parton model describes the experimental data well.
However, the predictions of the improved parton model clearly deviate from
the experimental data below $Q^2<$ 7 GeV$^2$ for Bjorken $x$
between 0.2 and 0.45.
This deviation is not a random fluctuation. It is present in a broad
range of $Q^2$ as will be shown in the next figure.
Similar deviations of the IPM from the experimental data can also be observed
for $F_2^p$ and $F_2^d$, albeit at smaller values of $Q^2$ \cite{SU1}.
This is consistent with the cancellation of higher-twists for $F_2^p$ or $F_2^d$
and with the lack of such a cancellation for $F_2^p - F_2^n$ as discussed
above. None of nuclear effects we know can explain this disagreement.
\footnote{
No strong $Q^2$-dependence is predicted for nuclear effects except of a
VDM-type shadowing \cite{BK_shadowing} for $x<$ 0.1.}
Moreover the agreement of the improved parton model with
experimental data for small values of $x$ is most probably accidental,
as shadowing effects which are not included in this calculation, but are
present in the data, would enhance the theoretical results by 0.005-0.015,
depending on $(x,Q^2)$ and the model used.
This would worsen the agreement of the IPM also for small values
of Bjorken $x <$ 0.1. For the sake of simplicity, we omit nuclear
corrections in the present paper. We have estimated them and found
that they do not affect our conclusions.
For comparison we show the prediction of our model\cite{SU1}
(thick solid line) with $Q_0^2$ = 0.79 GeV$^2$,
% and $\lambda_G$ = 0.575 GeV
%(a parameter of the Gaussian form factor $\Omega$ \cite{SU1})
with target mass corrections calculated according to \cite{GP76}.
The target mass effects become important at small $Q^2$ and larger $x$.
%They play also some role for the correct location of the maximum in
%$F_2^p-F_2^n$.

In order to better visualize the unexpected
within the IPM $Q^2$-dependence we present in Fig.3
$D_G$ as a funtion of $Q^2$ for some selected values of Bjorken
$x$ = 0.140, 0.275, 0.350.
This figure clearly demonstrates a failure of the IPM
as well as substantial negative higher-twists for $F_2^p - F_2^n$.
The trend of the experimental data is completely different to
the prediction of the standard IPM. Our model reproduces well the main trend
of the experimental data and provides a better description than the standard IPM.
However, the agreement of our simple model with experimental data is
not complete.
The structure function difference $F_2^p - F_2^n$ is
a relatively small quantity and therefore may be sensitive
to subtle effects.
At low $Q^2$, i.e. rather low energy for fixed Bjorken-$x$,
one should worry about a careful treatment of different
exclusive channels and their contribution to $F_2$.
One may also expect some contribution due to the $\rho - \omega$
nondiagonal VDM term (\ref{nondiagonal}).
NNLO analysis and careful treatment of nuclear effects could also
be valuable.

The effect discussed in the present paper may be very important
to understand small differences for light-antiquark flavour asymmetry
obtained from different types of experiments.
Recently the E866 experiment on Drell-Yan production in proton-proton and
proton-deuteron scaterring has provided a new, rather precise, information
about the nucleon $\bar d - \bar u$ asymmetry \cite{E866}. Because
of the broad range of $x$ available in this experiment we could
estimate
\begin{equation}
\int_0^1 dx \; [\bar d(x) - \bar u(x)]_{E866} = 0.09 \pm 0.02 \; .
\label{asymm_E866}
\end{equation}
This number is smaller than the one deduced from a somewhat
older NMC-result \cite{NMC} on the Gottfried Sum Rule
\footnote{We note that both analyses assume isospin symmetry
for proton and neutron quark distributions.}
\begin{equation}
\int_0^1 dx \; [\bar d(x) - \bar u(x)]_{NMC} = 0.147 \pm 0.039 \; .
\label{asymm_NMC}
\end{equation}
While the average photon virtuality in the Fermilab experiment was fairly
large $<Q^2> \sim$ 50 GeV$^2$, the average photon virtuality in the CERN
deep inelastic scattering experiment was relatively low: $<Q^2> \sim$
4 GeV$^2$. The latter experiment is in the range of $Q^2$ where we have
observed large negative higher-twist effects (see Figs.2 and 3). This
means that the deviation of the Gottfried Sum Rule from its classical value
of $\frac{1}{3}$ \cite{GSR} observed by NMC \cite{NMC} is partially
due to the $\bar d - \bar u$ asymmetry and partially due to the discussed
higher-twist effects as follows from Eq.(\ref{diff_part}).
The higher-twist effects would resolve then the discrepancy between
(\ref{asymm_E866}) and (\ref{asymm_NMC}).

In summary, we have found substantial deviations from the improved parton model
for the difference $F_2^p-F_2^n$ already at $Q^2$ as large as 7 GeV$^2$,
which is much higher than for $F_2^p$ and $F_2^d$ separately.
The deviation has been predicted by us in \cite{SU1}.
When combined, these two analyses strongly suggest that the agreement
of the improved parton
model for $F_2^p$ and $F_2^d$ at $Q^2 <$ 7 GeV$^2$ is rather accidental.
The present result is in contrast to rather small higher-twist effects
estimated in $F_3^{\nu N}(x,Q^2)$ \cite{F3}.
The nuclear effects in the deuteron have not been included
in the present analysis. However, we estimated their role within
simple nuclear models and found that taking into account these effects
would only strengthen our conclusions.
The results obtained here strongly suggest the $Q^2$-dependence
of the Gottfried Sum Rule and could resolve a discrepancy between
the $\bar u - \bar d$ asymmetry obtained in different processes.
The breaking of the parton model approach for the reported $Q^2$ range
would have important consequences for different
analyses of exclusive reactions concerning the flavour and spin
structure of the nucleon where the applicability of the parton model
is conditio sine qua non.

\newpage

\vskip 1cm

{\bf Acknowledgments}
We are indebted to Antje Br\"ull for the discussion of details of
the NMC data and Jan Kwieci\'nski for the discussion on
the Bade{\l}ek-Kwieci\'nski model.
We are also grateful to Rudolf Tegen for a careful reading of the manuscript
and useful comments. This work was partially supported
by the German-Polish DLR exchange program, grant number POL-028-98.

%------------------------------------------------------------------------

\newpage

%====================================================================

\begin{figure}
\mbox{
\epsfysize 15.0cm
\epsfbox{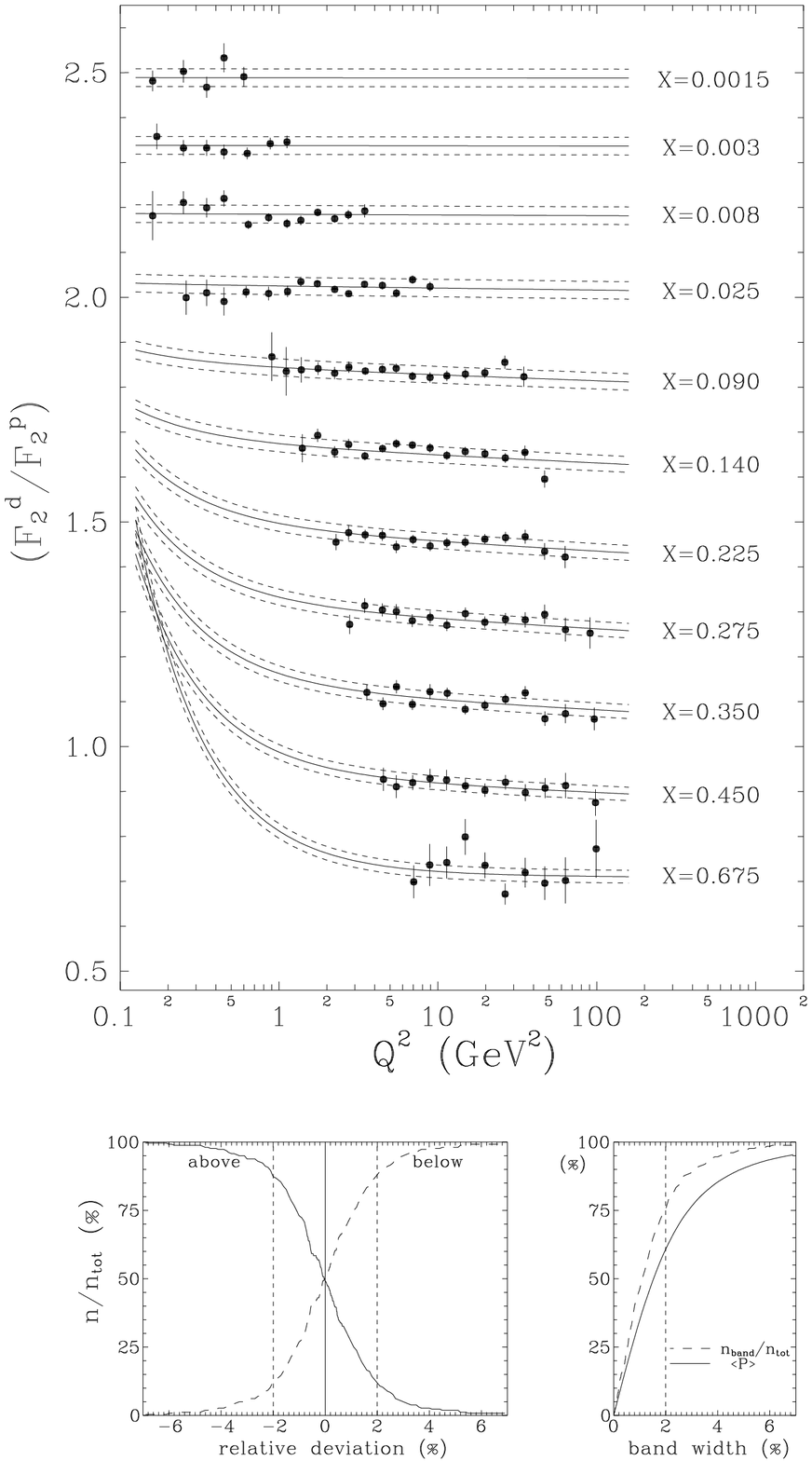}
}
\caption{\it
The quality of the $F_2^n/F_2^p$ ratio parametrization
for the NMC $F_2^d/F_2^p$ data \cite{NMC_97}.
In the top panel shown is $F_2^d/F_2^p$ shifted by 0, 0.15, ..., 1.5
for x = 0.675, 0.450, ..., 0.0015, respectively,
as a function of the photon virtuality.
The short-dashed line shows a 2 \% uncertainty of the ratio
as discussed in the text.
In the bottom-left panel shown is an up-down asymmetry of the NMC data
with respect to the used parametrization and
in the bottom-right panel
a fraction of the same data points in the parametrization uncertainty
band.
}
\label{F2_ratio}
\end{figure}

%====================================================================

\begin{figure}
\begin{center}
\mbox{
\epsfysize 15.0cm
\epsfbox{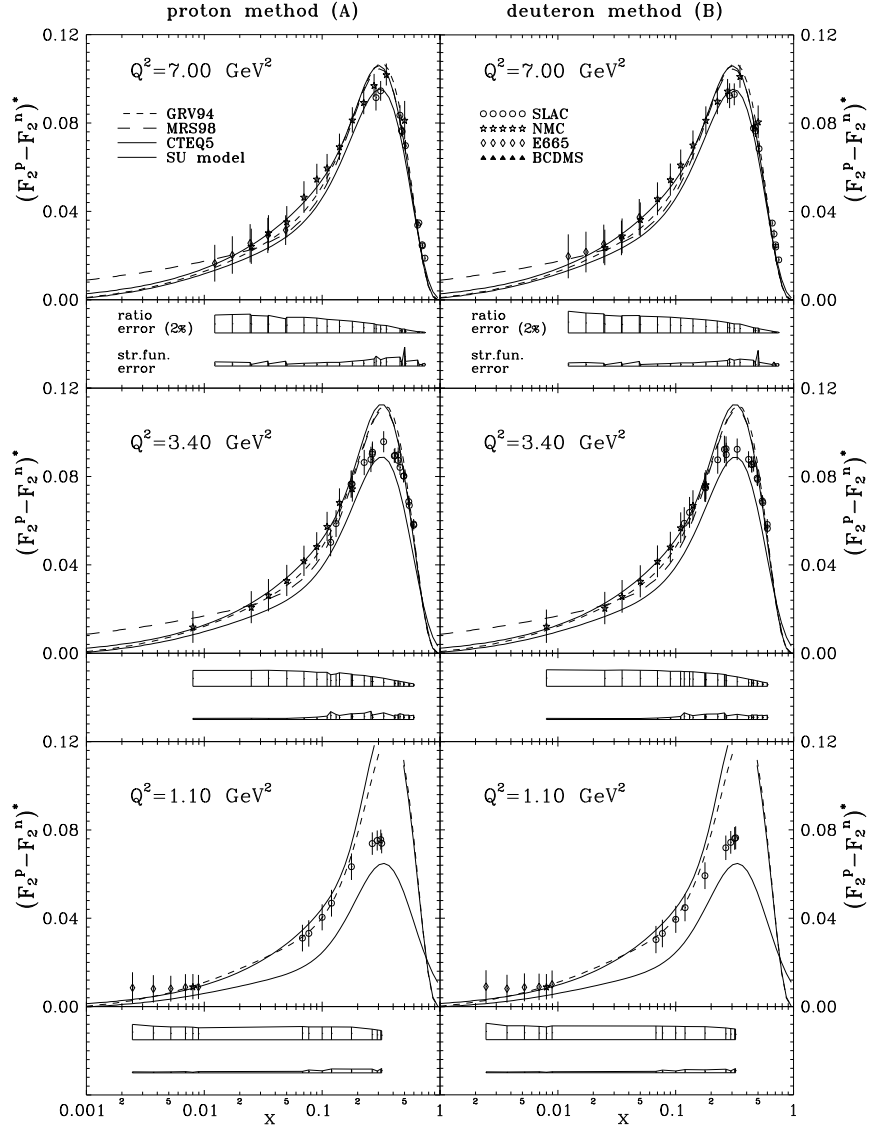}
}
\end{center}
\caption{\it
$D_G(x)$ for some selected $Q^2$ values obtained with
the method A (lhs) and the method B (rhs).
Below each distribution shown separately is an error band due to
the uncertainty of the $F_2^n / F_2^p$ ratio and an error band due
to the uncertainty of $F_2^p$ (lhs) or $F_2^d$ (rhs).
The short-dashed line correponds to the GRV94 NLO \cite{GRV94} PDF's,
the long-dashed line to the MRST98 NLO \cite{MSRT98} PDF's
and the solid to the CTEQ5 NLO \cite{CTEQ5} PDF's.
The thick solid line represents a prediction of our model \cite{SU1}
with details as described in the text.
}
\label{diff_x}
\end{figure}

%====================================================================

\begin{figure}
\begin{center}
\mbox{
\epsfysize 15.0cm
\epsfbox{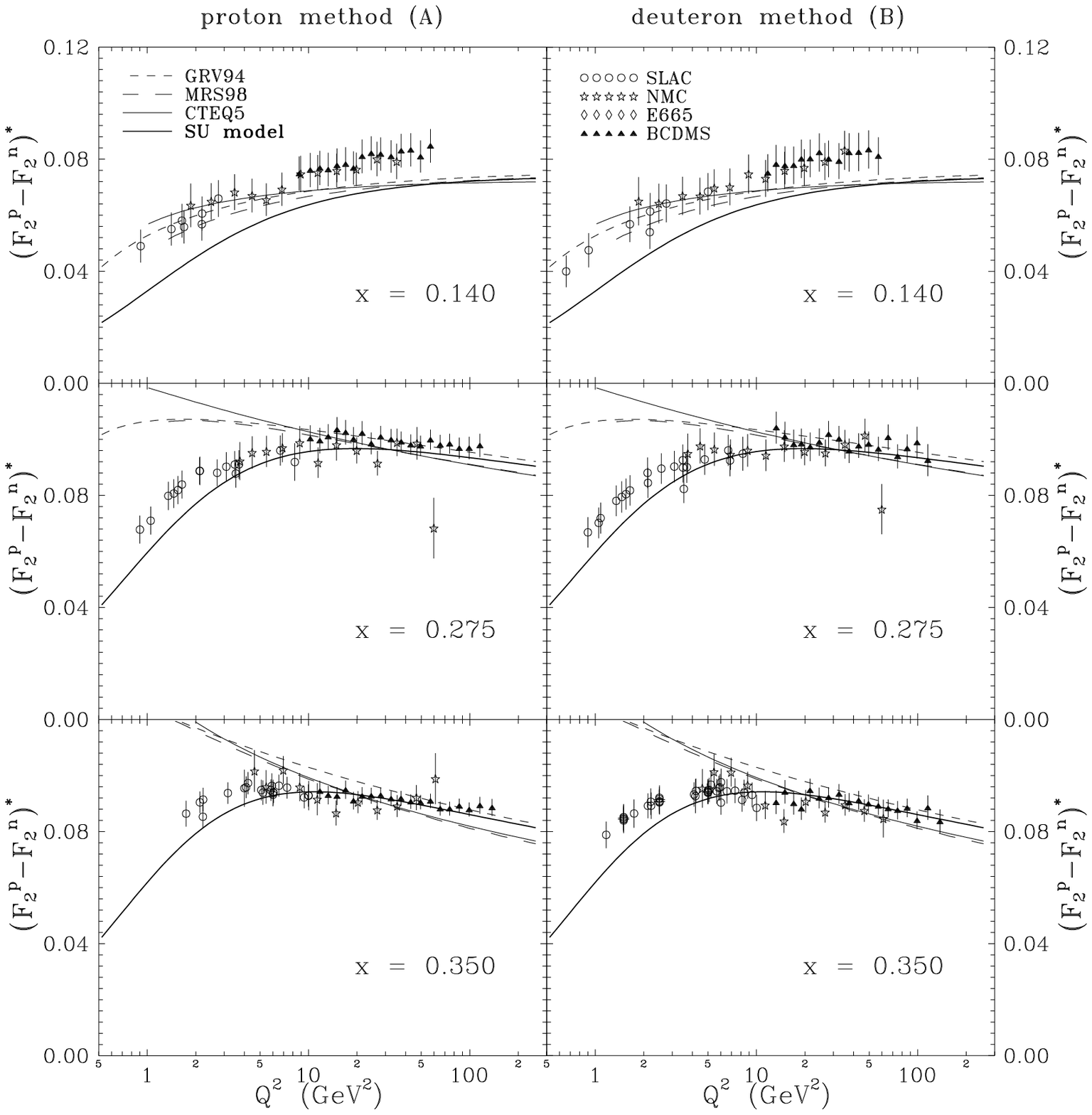}
}
\end{center}
\caption{\it
$D_G(Q^2)$ for some selected values of $x$.
The meaning of the curves is the same as in Fig.2.
}
\label{diff_Q2}
\end{figure}


\newpage

\begin{thebibliography}{99}

\bibitem{GRV94}
M. Gl\"uck, E. Reya and A. Vogt,
Z. Phys. {\bf C67} (1995) 433.

\bibitem{MSRT98}
A.D. Martin, R.G. Roberts, W.J. Sirling and R.S. Thorne,
hep-ph/9803445, Eur. Phys. Jour. {\bf C4} (1998) 463.

\bibitem{CTEQ5}
H.L. Lai et al.,
hep-ph/9903282.

\bibitem{VM92}
M. Virchaux and A. Milsztajn,
Phys. Lett. {\bf B274} (1992) 221.

\bibitem{NMC_ht}
NM collaboration, P. Amaudruz et al.,
Nucl. Phys. {\bf B371} (1992) 3.

\bibitem{lattice}
S. Capitani et al.,
hep-ph/9906320.

\bibitem{SU1}
A. Szczurek and V. Uleshchenko,
hep-ph/9904288, in print in Eur. Phys. Jour.C

\bibitem{BK_model}
J. Kwieci\'nski and B. Bade{\l}ek,
Z. Phys.{\bf C43} (1989) 251; \\
B. Bade{\l}ek and J. Kwieci\'nski,
Phys. Lett. {B295} (1992) 263.

\bibitem{BF94}
R.D. Ball and S. Forte,
Nucl. Phys. {\bf B425} (1994) 516.

\bibitem{NMC}
NM collaboration, M. Arneodo et al.,
Phys. Rev. {\bf D50} (1994) R1.

\bibitem{AB_habilitation}
A. Br\"ull, Habilitationsschrift, Heidelberg 1995.

\bibitem{NMC_97}
NM collaboration, M. Arneodo et al.,
Nucl. Phys. {\bf B487} (1997) 3.

\bibitem{E665_ratio}
E665 collaboration, M.R. Adams et al.,
Phys. Rev. Lett. {\bf 75} (1995) 1466.

\bibitem{F2_data}
http://durpdg.dur.ac.uk/HEPDATA.

\bibitem{BK_shadowing}
B. Bade{\l}ek and J. Kwieci\'nski,
Nucl. Phys. {\bf B370} (1992) 278.

\bibitem{GP76}
H. Georgi and H.D. Politzer,
Phys. Rev. {\bf 14} (1976) 1829.

\bibitem{E866}
E866/NuSea collaboration, E.A. Hawker et al.,
Phys. Rev. Lett. {\bf 80} (1998) 3715.

\bibitem{GSR}
K. Gottfried, Phys. Rev. Lett. {\bf 18} (1967) 1174.

\bibitem{F3}
A.L. Kataev, A.V. Kotikov, G. Parente and A.V. Sidorov,
Phys. Lett. {\bf B417} (1998) 374.

\end{thebibliography}
\end{document}